\documentclass[12Pt]{article} 
\usepackage{amssymb,euscript,latexsym,amsmath} 
\usepackage{graphicx} 
\usepackage{epstopdf} 
\usepackage{bm} 
\usepackage{enumerate} 
\usepackage{setspace}
\usepackage{subfigure}
 \usepackage{hyperref}

\begin{document}

\title{Motion Control of a Spinning Disc on Rotating Earth} 
\author{Fangxu Jing, Eva Kanso and Paul K. Newton}
\maketitle 

\begin{abstract} 
	This paper considers the motion control of a particle and a spinning disc on rotating earth.
The equations of motion are derived using Lagrangian mechanics. Trajectory planning is studied as an optimization problem
using the method referred to as Discrete Mechanics and Optimal Control. 
\end{abstract}

\section{Introduction}

This paper considers the motion control of a particle and a spinning disc on earth. In particular, we derive the governing
equations using Lagrangian mechanics and study trajectory planning as an optimization problem using the method referred to
as Discrete Mechanics and Optimal Control, see~\cite{JuMaOb2005, KaMa2005}.

\begin{figure}[!h] 
	\centering 
	\includegraphics[width=2.75in]{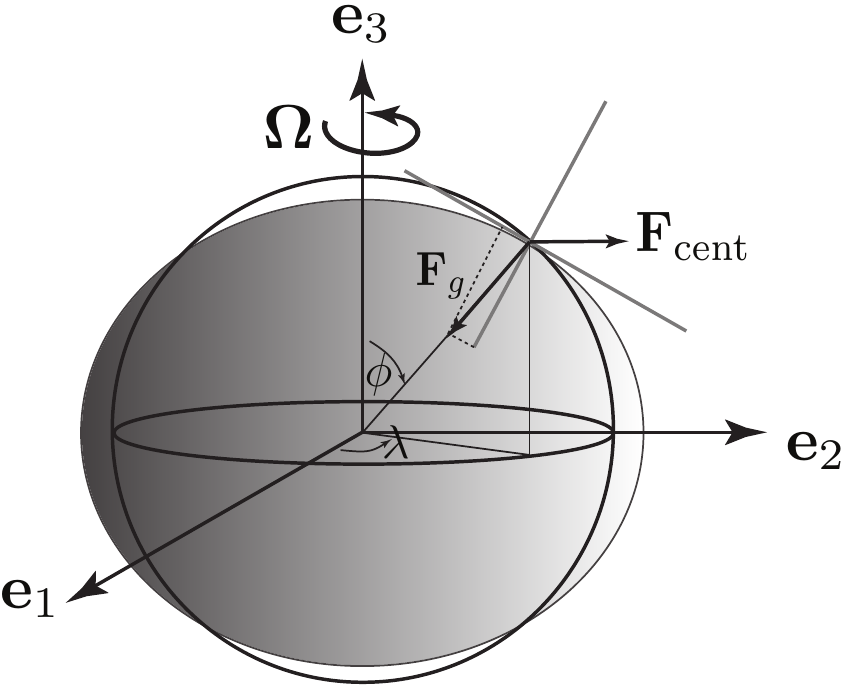}
\caption{\footnotesize The earth (oblate spheroid) is modeled as a rotating sphere with an added potential used to emulate
the effect of the eccentricity which produces a non-zero horizontal component of gravity. } \label{fig:model} 
\end{figure}

\subsection{Problem Description} 
Consider the horizontal motion on earth in the absence of friction and pressure gradient
forces. A perfect sphere model fails to capture the dynamical effects associated with the horizontal component of gravity
due to the eccentricity of the earth, as shown in Figure~\ref{fig:model} and discussed
in~\cite{Coushman1982, PaKi1988, PaSi2001, Paldor2005, Ripa1997} for a particle moving on
rotating earth. It also plays an important role in the unsteady dynamics of a spinning disc on rotating earth as
investigated in~\cite{McDonald1998, Ripa2000_1, Ripa2000_2} as a model for geophysical vortex motion. In
both problems, the authors emulated the effect of the earth eccentricity by adding an artificial potential to the perfect
sphere model. In this paper, we use the same model to address the problem of planning the motion on earth as an
optimization problem based on a Lagrangian formulation. More specifically, we investigate {\em optimal trajectories} that
steer the particle or the spinning disc from an initial position and velocity to a final position and velocity while
minimizing a prescribed cost function such as the control effort.

\subsection{Motivation} 
The main motivation for these problems comes from atmospheric sciences where one needs to track
atmospheric drifters over periods of days, weeks, or even months. In the simplest case, these drifters are high altitude
balloons which, to a first order approximation, can be thought of as passive tracers. However, if complex dynamical
maneuvers are required, it is useful to include more complicated internal dynamics, such as rotational motion, which adds
interesting new effects. These kinds of finite-dimensional models have also been used as mechanical models of vortex motion
in some geophysical settings \cite{McDonald1998} and one could imagine thinking of the spinning disc model, in the limit as
the ratio of the disc radius to the radius of the sphere goes to zero, as a ``point-vortex'', a collection of which might
approximate a distribution of vorticity~\cite{Newton2001}. Interestingly, systems of interacting rotating millimeter-sized
discs floating on a liquid-air interface have been used recently~\cite{GrStWh2000} to study self-assembly of patterns in
models that have some of the same pattern forming features as a wide range of vortex lattice systems. We consider the
optimal control problems treated in this paper as a first step in the process of attempting to implement control and motion
strategies in these contexts.

\subsection{Organization of the Paper} In~Section~\ref{sec:formulation}, we formulate the dynamics of a particle and a
spinning disc moving on earth, and address the question of motion planning in~Section~\ref{sec:control}. The numerical
implementation and discussion of the results are presented in~Section~\ref{sec:num}.

\section{Dynamics on Rotating Earth} 
\label{sec:formulation}

Consider a particle of mass $m$ moving on a sphere of radius $R$ and subject only to a gravitational force pointing towards
the center of the sphere and a supporting force normal to the surface of the sphere (i.e., pointing in the opposite
direction of gravity). One distinguishes two types of behavior relative to a fixed inertial frame
$\{\mathbf{E}_i\}_{i=1,2,3}$ as follows: if the initial velocity is zero, the particle remains at rest for all time,
otherwise, it moves along a great circle which passes through the starting position and is tangent to the direction of the
initial velocity at the starting point. If the sphere is rotating, say with an angular velocity vector
$\boldsymbol{\Omega}= \Omega \, \mathbf{E}_3$ where $\Omega=2\pi/1$day, the motion of the particle can be described
relative to either a fixed inertial frame or a frame rotating with the sphere $\{\mathbf{e}_i\}_{i=1,2,3}$ with
$\mathbf{e}_3=\mathbf{E}_3.$ In the inertial frame, one sees the same types of trajectories as in the stationary sphere
case. The motion appears more complicated in the rotating frame where Coriolis and centrifugal forces need to be taken into
consideration. An important observation here is that a particle starting at rest relative to the moving frame does not
remain at rest due to the {\em horizontal component} of the centrifugal force, i.e., the component in the direction tangent
to the earth surface. However, a particle starting at rest relative to the rotating earth remains at rest for all time.
Indeed, in a more accurate model of the earth where the effect of eccentricity is taken into account, the horizontal
component of the centrifugal force $\mathbf{F}_{\rm cent}$ is counterbalanced by the horizontal component of the
gravitational force $\mathbf{F}_g$. Recall that $\mathbf{F}_{\rm cent}$ is given by: 
\begin{equation}
\mathbf{F}_{\rm cent} = -m\, \boldsymbol{\Omega}\times(\boldsymbol{\Omega}\times \mathbf{R}), 
\end{equation}
where $\mathbf{R}$ denotes the position vector of the particle on the sphere. It is convenient to introduce
spherical coordinates $(\phi,\lambda)$ measured relative to the moving frame $\{\mathbf{e}_i\}$,
see~Figure~\ref{fig:model}, and an associated right-handed orthonormal basis $\{\mathbf{e}_{\phi},\mathbf{e}_{\lambda},
\mathbf{e}_{R}\}.$ The centrifugal force $\mathbf{F}_{\rm cent}$ can then be expressed as follows: 
\begin{equation} 
	\mathbf{F}_{\rm cent} = mR\,\Omega^{2}\sin\phi \, ( \sin\phi \, \mathbf{e}_{R} + \cos\phi \,
\mathbf{e}_{\phi}). 
\end{equation} 
To cancel the horizontal component of $\mathbf{F}_{\rm cent}$, the
particle should be subject to a horizontal force opposite in direction and of magnitude equal to
$mR\Omega^{2}\sin\phi\cos\phi$. One way to incorporate this effect yet retain the obvious advantages of spherical
coordinates with constant radius is to introduce a potential function $V_a$ whose gradient in the horizontal direction
produces such force,~\cite{Ripa1997}, namely, 
\begin{equation}\label{eq:Va} 
	V_{a} =
\frac{1}{2}\,mR^{2}\,\Omega^{2}\,\sin^{2}\phi . 
\end{equation}

\subsection{Equations of Motion for a Particle}

The Lagrangian function for the system is given by: 
\begin{equation}\label{eq:Lag} 
	L(q,\dot{q}) = T - V ,
\end{equation} 
where $T$ and $V$ denote the kinetic and potential energies, respectively, while $q$
parameterizes the position of the particle on earth and can be chosen as $q = (\phi,\lambda)$. Lagrange's equations of
motion are given by {\em Hamilton's principle} (also known as the {\em least action principle}). This principle amounts to
taking variations of the action, between a fixed initial time $t_0$ and a fixed final time $t_f$,
\begin{equation} 
	S = \int_{t_0}^{t_f} L \,{\rm d}t 
\end{equation} 
with respect to arbitrary variations
$\delta q$ of the path that keep the endpoints $q(t_0)$ and $q(t_f)$ fixed. The point-wise equations of motion thus obtained
are of the form: 
\begin{equation}\label{eq:Lageom} 
	\dfrac{{\rm d}}{{\rm d}t}\dfrac{\partial L}{\partial \dot{q}}
- \dfrac{\partial L}{\partial {q}} = 0 . 
\end{equation}
The kinetic energy $T$ of the particle can be expressed in terms of spherical coordinates as: 
\begin{equation}\label{eq:Tparticle} 
	T = \frac{1}{2}\,mR^{2}\left[{\dot{\phi}}^{2}+(\dot{\lambda}+\Omega)^{2}\sin^{2}\phi\right]. 
\end{equation}
The potential energy is the sum of two parts, the gravitational potential $V_g$ which is a constant and therefore can be set to zero, and the artificial potential $V_a$ introduced in~\eqref{eq:Va}. Lagrange's equations~\eqref{eq:Lageom} read as:
\begin{equation}\label{eq:eomparticle} 
	\begin{cases}
\ddot{\phi}-\dot{\lambda}\Omega\sin2\phi-\dfrac{1}{2}\,{\dot{\lambda}}^{2}\sin2\phi = 0,\\[1ex]
 \ddot{\lambda}+2\,
\dot{\phi}\,(\dot{\lambda}+\Omega)\cot\phi = 0 . 
\end{cases} 
\end{equation}
Finally, note that, since $L$ is not an explicit function of longitude $\lambda$ and time $t$, by Noether's theorem, one has two integrals of motion (conserved quantities) associated with these symmetries. The angular momentum in the $\mathbf{e}_\lambda$-direction is
conserved, 
\begin{equation*} 
	\pi_{\lambda} = mR^{2}(\dot{\lambda}+\Omega)\sin^{2}\phi = {\rm
constant} , 
\end{equation*}
and the total energy $ E$ is conserved 
\begin{equation*}
	E = \frac{1}{2}mR^{2}\left[{\dot{\phi}}^{2}+(\dot{\lambda}+\Omega)^{2}\sin^{2}\phi\right]+
\frac{1}{2}mR^{2}\Omega^{2}\sin^{2}\phi = {\rm constant}. 
\end{equation*}

\begin{figure*}[!t] 
	\centerline{ \includegraphics[width=1.15\textwidth]{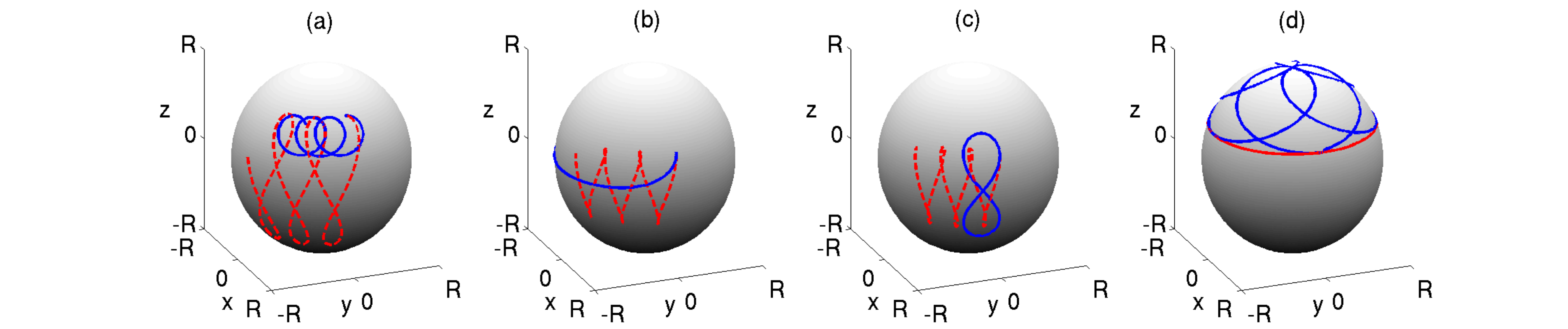}}
\caption{\footnotesize Free motion of a particle of mass $m$ on a rotating earth and rotating sphere, the solid line is the
trajectory on earth, dashed line is on sphere. End time is 3 days for all examples, initial conditions are: (a)
$(\phi_0,\lambda_0)=(\pi/4,0)$, $(\dot{\phi}_0,\dot{\lambda}_0)=(0,2)$; (b) $(\phi_0,\lambda_0)= (3\pi/8,0)$,
$(\dot{\phi}_0,\dot{\lambda}_0)=(0.92,0)$; (c) $(\phi_0,\lambda_0)=(3\pi/8,0)$, $(\dot{\phi}_0,\dot{\lambda}_0)=(1.366,0)$;
(d) $(\phi_0,\lambda_0)= (3\pi/8,0)$, $(\dot{\phi}_0,\dot{\lambda}_0)=(0,-2\pi)$.} \label{fig:numintparticle}
\end{figure*}

\subsection{Equations of Motion for a Spinning Disc}
\label{sec:eomdisc}

Consider a thin disc, of radius $a$ and uniformly distributed mass $m$, tangent to the earth surface such that its center
of mass lies on earth and can be parameterized by $(\phi,\lambda)$. The disc is free to rotate or spin about its axis of
symmetry which, by assumption, remains normal to the earth surface at all time. That is, the disc is free to spin about the
$\mathbf{e}_R$-axis.

The Lagrangian function of the spinning disc is given by~\eqref{eq:Lag}. The kinetic energy can be decomposed into a
translation kinetic energy which has the same form as in~\eqref{eq:Tparticle} and a rotational kinetic energy $T_{\rm
rot}$, which can be expressed as: 
\begin{equation}\label{eq:Tdisc} 
	T_{\rm rot} =
\frac{1}{2}I_{h}\,\omega_{\phi}^{2} +\frac{1}{2}I_{h}\,\omega_{\lambda}^{2}+\frac{1}{2}I_{v}\,\omega_{\psi}^{2} ,
\end{equation} 
where $I_{v}=\frac{1}{2}ma^{2}$ and $I_{h}=\frac{1}{4}ma^{2}$ are the moments of inertia
of the disc about $\mathbf{e}_R$ and about the horizontal axes $\mathbf{e}_\phi$ and $\mathbf{e}_\lambda$, respectively.
In~\eqref{eq:Tdisc}, the components of the angular velocity relative to the spherical basis
$\{\mathbf{e}_{\phi},\mathbf{e}_\lambda,\mathbf{e}_R\}$ are given by: 
\begin{equation}\label{eq:angvel}
\omega_{\phi}=\dot{\phi},\quad \omega_{\lambda}=(\dot{\lambda}+\Omega)\sin\phi ,\quad
\omega_{\psi}=\dot{\psi}+(\dot{\lambda}+\Omega)\cos\phi . 
\end{equation}
Now, use $M=mR^{2}$,
substitute~\eqref{eq:angvel} in~\eqref{eq:Tdisc} and add~\eqref{eq:Tparticle} to get the total kinetic energy
\begin{equation} 
T =
\frac{1}{2}\,(M+I_{h})\left[{\dot{\phi}}^{2}+(\dot{\lambda}+\Omega)^{2}\sin^{2}\phi\right]+
\frac{1}{2}\,I_{v}\left[\dot{\psi}+(\dot{\lambda}+\Omega)\cos\phi\right]^{2}. 
\end{equation}

To simplify the expression for the potential energy, we assume that $a\ll R$ and treat the gravitational and artificial
potentials, $V_g$ and $V_a$, as functions of the center of mass only. In addition to $V_g$ and $V_a$, the disc is subject
to a potential function $V_d$ of the form
\begin{equation} 
	V_{d} =
\frac{1}{2}\,I_{v}\,\Omega^{2}\cos^{2}\phi , 
\end{equation}
which, as $V_a$, is due to the fact that the
dynamics is described in a rotating frame. To better understand the potential $V_d$, consider the special case when the
disc is stationary in inertial frame and its center of mass is placed at the north pole $\phi=0$. Clearly, relative to the
rotating frame, the disc is not stationary and has total energy $\frac{1}{2}I_{v}\Omega^{2}$, which is exactly
$V_{d}|_{\phi=0}$. The total potential energy $V = V_g + V_a + V_d$ can then be expressed as
\begin{equation} 
	V \ = \ \frac{1}{2}\,\Omega^{2}(M\sin^{2}\phi+I_{v}\cos^{2}\phi), 
\end{equation}
where, as before, $V_g$ is constant and is set equal to zero. Lagrange's equations~\eqref{eq:Lageom} for the spinning disc, with
$q=(\phi,\lambda,\psi)$ and $I_v=2I_h=2I$, yield
\begin{equation}
	\label{eq:eomdisc} 
	\begin{cases}
(M+I)\ddot{\phi} = - 2I\dot{\psi}(\dot{\lambda} + \Omega)\sin\phi + \dfrac{1}{2}\left[(M-I)(\dot{\lambda}+\Omega)^{2} -(M-2I)\Omega^{2}\right]\sin2\phi,\\[1ex]
(M+I)\ddot{\lambda}\sin\phi = 2I\dot{\psi}\dot{\phi} -2M(\dot{\lambda}+\Omega)\dot{\phi}\cos\phi,\\[1ex]
(M+I)\ddot{\psi}\sin\phi = -2I\dot{\psi}\dot{\phi}\cos\phi + \left[M(1+\cos^2\phi)+I\sin^2\phi\right](\dot{\lambda}+\Omega)\dot{\phi}. 
\end{cases} 
\end{equation}
The Lagrangian function $L$ does not depend explicitly on $\lambda$, $\psi$ and time $t$ which
implies the existence of three integrals of motion. The angular momenta $\pi_\lambda$ and $\pi_\psi$ in the
$\mathbf{e}_\lambda$ and $\mathbf{e}_R$ directions are conserved,
\begin{equation*} 
	\begin{cases}
\pi_{\lambda} = (M+I)(\dot{\lambda}+\Omega)\sin^{2}\phi + 
2I\left[\dot{\psi}\cos\phi+(\dot{\lambda}+\Omega)\cos^{2}\phi\right] = {\rm constant}, \\[1ex] 
\pi_{\psi} = I_{v}\left[\dot{\psi}+(\dot{\lambda}+\Omega)\cos\phi\right] = {\rm constant} , 
\end{cases} 
\end{equation*}
as well as the total energy
\begin{equation*} 
		E = 
\frac{1}{2}M\Omega^{2}\sin^{2}\phi+\frac{1}{2}(M+I)\left[{\dot{\phi}}^{2}+ (\dot{\lambda}+\Omega)^{2}\sin^{2}\phi\right]+
I\left\{ \left[\dot{\psi}+(\dot{\lambda}+\Omega)\cos\phi\right]^{2} +\Omega^{2}\cos^{2}\phi\right\}=  {\rm
constant} . 
\end{equation*} 
An approximate solution to~\eqref{eq:eomdisc} can be sought under
the assumption that the disc translational velocity is much smaller than the velocity of the earth, i.e.
$\dot{\phi}\ll\Omega$ and $\dot{\lambda}\ll\Omega$. The approximate solution, referred to as ``steady drift''
in~\cite{McDonald1998}, is: 
\begin{equation} 
	\phi\approx\phi_{0},\quad
\dot{\psi}\approx\dot{\psi}_{0},\quad \dot{\lambda}\approx\frac{1}{2}\frac{I_{v}}{M}\frac{\dot{\psi}_{0}}{\cos\phi_{0}}.
\end{equation}
The comparison between this approximate solution and our direct numerical integration
of~\eqref{eq:eomdisc} suggests that this is a good approximation under additional conditions (see Section~\ref{sec:num} for
more details). Namely, $\dot{\psi}$ has to be much larger than $\Omega$, say of the order $O(10\Omega)$. Also, it should be
clear that the latitude $\phi$ must not be chosen close to the equator where $\dot{\lambda}$ obtained above is not accurate
and the motions in $\phi$ and $\psi$ are too significant to ignore.

\begin{figure*}[!t] 
	\centerline{ \includegraphics[width=1.1\linewidth]{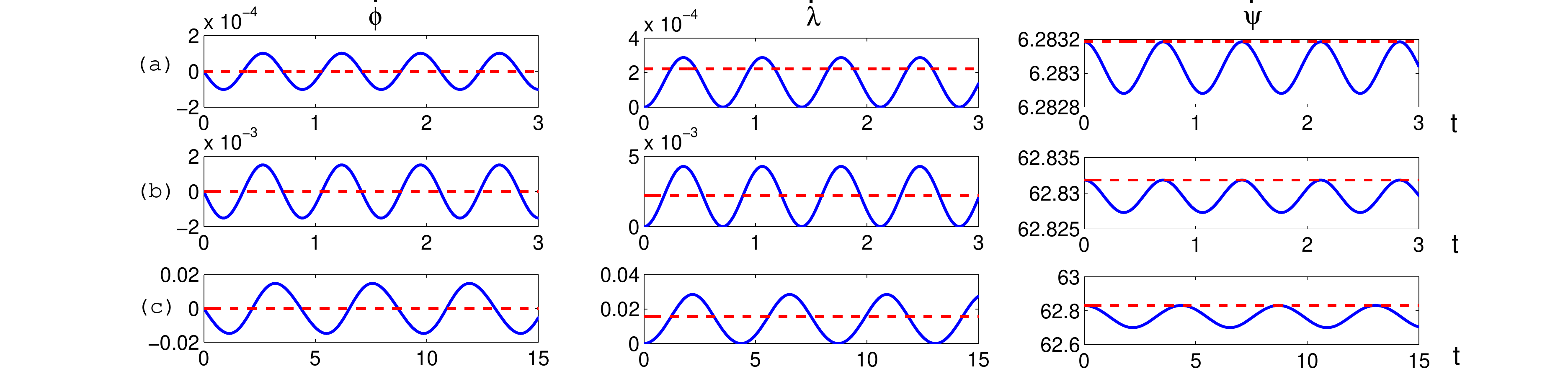}}
\caption{\footnotesize Free motion of a spinning disc on rotating earth, the solid line is the result of direct numerical
integration, dashed line is an approximate solution. We present $\dot{\phi}$, $\dot{\lambda}$ and $\dot{\psi}$ in three
columns respectively. Initial conditions and end times are: (a) $(\phi_0,\lambda_0)=(\pi/4,0)$,
$(\dot{\phi}_0,\dot{\lambda}_0,\dot{\psi}_0)=(0,0,\Omega)$, end time 3 days; (b) $(\phi_0,\lambda_0)=(\pi/4,0)$,
$(\dot{\phi}_0,\dot{\lambda}_0,\dot{\psi}_0)=(0,0,10\Omega)$, end time 3 days; (c) $(\phi_0,\lambda_0)=(\pi/2-0.1,0)$,
$(\dot{\phi}_0,\dot{\lambda}_0,\dot{\psi}_0)=(0,0,10\Omega)$, end time 15 days.} \label{fig:numintdisc}
\end{figure*}

\section{Motion Planning} 
\label{sec:control}

The main point of this article is to investigate the problem of motion planning for the particle and the spinning disc
moving on earth. In the case of the particle, we introduce control forces in both $\mathbf{e}_\phi$ and
$\mathbf{e}_\lambda$ direction and solve for optimal control forces that steer the particle from an initial position and
velocity to a desired final position and velocity. For the spinning disc, we apply a control torque about the axis of
symmetry of the disc, i.e., a torque that controls the spin $\psi$ of the disc. Clearly, this spinning disc is
under-actuated and questions of solvability and controllability are important.

\subsection{Solvability and Controllability} 

The problem for the spinning disc is to find optimal torques that steer
$\bigl(\phi(t_0),\lambda(t_0),\lambda(t_0)\bigr)$ to $\bigl(\phi(t_f),\lambda(t_f),\lambda(t_f)\bigr).$ For some given
initial and final conditions, there may be no control torque that achieves the desired motion. In this case the
optimization problem has no solution.

In Section~\ref{sec:num}, we provide numerical evidence that the spinning disc can achieve a net motion on the earth
surface under control mechanism. This suggests that the problem may be controllable or, at least, controllable in some
finite regions of the earth surface. For a rigorous proof of controllability, one needs to appeal to controllability
theorems for systems with drift, see, e.g.,~\cite{NiSc1990}. Such undertaken, although very important, is beyond the scope
of the present paper.

\subsection{Motion Control as an Optimization Problem} 

Take $q $ to be the state variables and let $f$ be the control
force. For the case of a particle, one has $q= (\phi,\lambda)$ and $f = (f_\phi,f_\lambda)$ while for the spinning disc $q=
(\phi,\lambda,\psi)$ and $f = (0,0,f_\psi)$. The motion planning problem can then be stated as follows. Given the boundary
conditions $q(t_0)=q_0$, $\dot{q}(t_0)=\dot{q}_0$ and $q(t_f)=q_f$, $\dot{q}(t_f)=\dot{q}_f$, find $f$ that minimizes the
cost function
\begin{equation}\label{eq:cost} 
	\int_{t_0}^{t_f} C(q,\dot{q}, f) \, \rm{d} t
\end{equation}
subject to the Lagrange-D'Alembert principle 
\begin{equation} 
	\label{eq:restate} 
	\delta
\int_{t_0}^{t_f} L(q,\dot{q}) \, {\rm d}t \,+ \int_{t_0}^{t_f} f \cdot \delta q \, {\rm d}t \, +\, p_0 \cdot \delta q_0 \,
- \, p_f \cdot \delta q_f = 0 , 
\end{equation} 
for all arbitrary variations $\delta q$. That is, the least action
principle outlined in Section~\ref{sec:formulation} is restated here as the Lagrange-d'Alembert principle (to account for
external control forces) and without the a priori assumption that the variations vanish at the end points $t_0$ and $t_f$.
Rather, this condition is imposed using the boundary constraints 
\begin{equation}
\label{eq:constraints} 
\delta {q}_0 = q(t_0) - q_0 = 0 \ , \quad \delta q_f = q(t_f) - q_f = 0 ,
\end{equation}
and their associated Lagrange multipliers 
\begin{equation}
\left. p_0 = \frac{\partial L}{\partial \dot{q}} \right|_{t_0}, \quad \left. p_f = \frac{\partial L}{\partial
\dot{q}}\right|_{t_f} . 
\end{equation}

\begin{figure*}[!t] 
	\centerline{\includegraphics[width=1.1\textwidth]{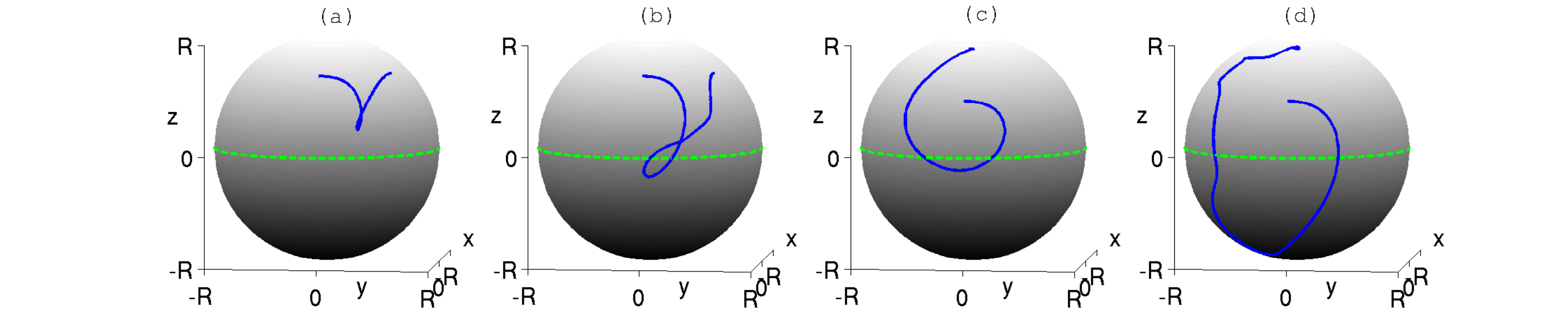}} 
	\caption{\footnotesize Optimal control results of
particle on rotating earth, the solid line is trajectory under control, and dashed line is equator. Initial and final
conditions for case (a) and (b) are: $(\phi_0,\lambda_0)=(\pi/4,0)$, $(\dot{\phi}_0,\dot{\lambda}_0)=(0,0)$,
$(\phi_f,\lambda_f)= (\pi/4,\pi/3)$, $(\dot{\phi}_f,\dot{\lambda}_f)=(0,0)$, and the end times: (a) $t_f=$1 day, (b)
$t_f$=3 days. Initial and final conditions for case (c) and (d) are: $(\phi_0,\lambda_0)=(\pi/3,0)$,
$(\dot{\phi}_0,\dot{\lambda}_0)=(0,0)$, $(\phi_f,\lambda_f)= (\pi/8,0)$, $(\dot{\phi}_f,\dot{\lambda}_f)=(0,0)$, and the
end times: (c) $t_f=$1 day, (d) $t_f$=3 days.} \label{fig:controlparticle}
\end{figure*}

\subsection{Discretization} 

We discretize the optimal control problem using the novel method devised by~\cite{JuMaOb2005}
where the idea is to discretize the cost function~\eqref{eq:cost} and the variational principle~\eqref{eq:restate} directly
using global discretization of the states and the controls. To this end, a path $q(t)$, where $t \in [t_0,t_f]$, is
replaced by a {\em discrete path} $q_d: \{0, h, 2h, \ldots, Nh=t_f]$, $N\in \mathbb{N}$. Here, $q_d(nh) := q_n$ is viewed
as an approximation to $q(t_n = nh)$, $n \in \mathbb{N}$ and $n \leq N$. Similarly, the continuous force $f$ is
approximated by a discrete force $f_d$ such that $f_{ n} = f_{ d}(n h)$.

The cost function~\eqref{eq:cost} is approximated on each time interval $[nh,(n+1)h]$ by
\begin{equation} 
	C_d(q_n, q_{n+1}, f_{ n}, f_ {n+1}) \approx \int_{nh}^{(n+1)h} C(q,\dot{q},f)\, {\rm d} t ,
\end{equation} 
which yields the discrete cost function 
\begin{equation}
\label{eq:discrete_cost} 
J_d(q_d, f_d) = \sum_{n=0}^{N-1} C_d(q_n, q_{n+1}, f_{ n}, f_{n+1}) . 
\end{equation}
The action integral~\eqref{eq:restate} is approximated on each time interval $[nh,(n+1)h]$ by a {\em
discrete Lagrangian} 
\begin{equation} 
	L_d(q_n, q_{n+1}) \approx \int_{nh}^{(n+1)h} L(q,\dot{q})
\,{\rm d} t . 
\end{equation}
We also approximate 
$\int_{t_n}^{t_{n+1}} f \cdot \delta q \approx f_{ n}^- \cdot \delta q_{ n} + f_{ n}^+ \cdot \delta f_{ n+1}$,
where $f^-_{ n}$ and $f^+_{ n}$ are called left and right discrete torques,
respectively. The discrete version of~\eqref{eq:restate} requires one to find paths $\{q_n\}_{n=0}^N$ such that for all
variations $\{\delta q_n\}_{n=0}^N$, one has 
\begin{equation}
	\label{eq:discrete_var} 
\delta \sum_{n=0}^{N-1} L_d(q_n, q_{n+1}) + \sum_{n=0}^{N-1} (f_{n}^- \cdot \delta q_{n} + f_{ n}^+ \cdot \delta q_{n+1})
+ p_0 \cdot \delta q_0 - p_f \cdot \delta q_f = 0. 
\end{equation}
The discrete
variational principle~\eqref{eq:discrete_var} yields the following equality constraints
\begin{equation}
	\label{eq:discrete_eq} 
	\begin{cases} 
		D_2 L_d (q_{n-1},q_n) + D_1 L_d (q_{n},q_{n+1}) + f_{n-1}^+ + f_{n}^-
 = 0 , \\[1ex]
  p_0 + D_1 L_d(q_0,q_1) + f_0^-  = 0 , \\[1ex]
   - p_f + D_2 L_d(q_{N-1},q_N) + f_{N}^+  = 0 . 
\end{cases}
\end{equation}

\section{Implementation and Numerical Results} 
\label{sec:num}

We now present some numerical results for free motion and motion planning of a particle and a spinning disc on earth.

\subsection{Direct Numerical Integration} 

The free motion of a particle on rotating earth is shown in
Figure~\ref{fig:numintparticle}. In order to illustrate the effect of the artificial potential $V_a$, we
integrated~\eqref{eq:eomparticle} for $V_a =0$ and for $V_a$ given in~\eqref{eq:Va} and plotted trajectories for the same
initial conditions in Figure~\ref{fig:numintparticle}. The solid line corresponds to trajectories on earth while the dashed
line is for trajectories on the sphere ($V_a =0$). The end time is 3 days in all examples. One can observe that due to the
existence of $V_a$, with some initial conditions (Figure~\ref{fig:numintparticle}(a)) the particle doesn't have enough
momentum in the latitudinal direction to overcome the peak of $V_{a}$ on the equator, therefore the trajectories are
constrained within one hemisphere, while the trajectories on the sphere always cross the equator when $\dot{\phi}_0\ne 0$.
In Figure~\ref{fig:numintparticle}(b), the trajectories asymptotically approach the equator. In
Figure~\ref{fig:numintparticle}(c), the trajectories appear to have an ``8'' shape and the cross point is always on the
equator (also mentioned in Paldor and Sigalov~\cite{PaSi2001}), note this is also observed in the sphere model but with
different initial conditions. Note that, since the motion on earth is actually motion on a sphere under a potential $V_a$,
the trajectories are similar to those of a spherical pendulum, see Figure~\ref{fig:numintparticle}(d). Here, for the same
initial conditions, the particle under no potential ($V_a = 0$) moves along a latitudinal circle.

Equations~\eqref{eq:eomdisc} for the free motion of a spinning disc on earth are integrated numerically and the results are
shown in~Figure~\ref{fig:numintdisc}. All examples are chosen to have zero initial translational velocity, which in the
particle case will lead to a stationary state for all time. By comparing all three sets of initial conditions, we affirm
our conclusions in~Section~\ref{sec:eomdisc} on the validity of the approximate solution. Namely, if $\dot{\psi}$ is too
small (Figure~\ref{fig:numintdisc}(a)), the ``steady drift'' velocity is not as good an average of $\dot{\lambda}$ as the
case where $\dot{\psi}$ is larger (Figure~\ref{fig:numintdisc}(b)). If the motion takes place near the equator
(Figure~\ref{fig:numintdisc}(c)), oscillations in all directions become too large to be ignored.

\begin{figure*}[!t] 
	\centerline{ \includegraphics[width=1.1\textwidth]{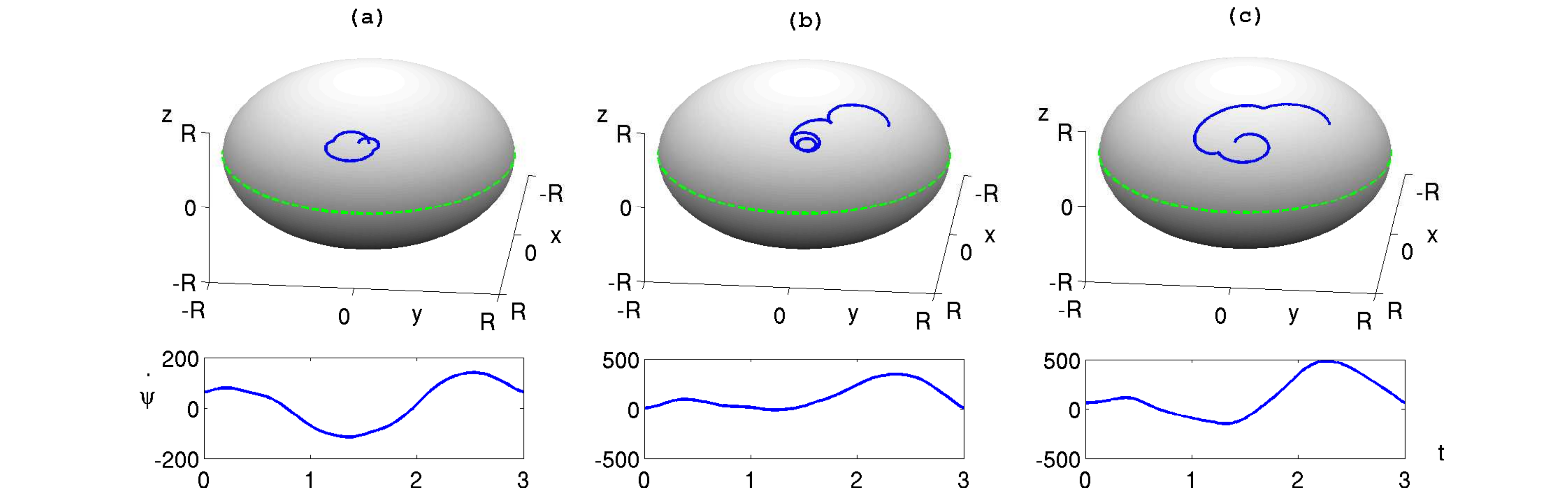}}
\caption{\footnotesize Optimal control results of the spinning disc on rotating earth. In the first row, we show
trajectories in solid line and the equator in dotted line. In the second row, we present the spinning velocities
correspondingly. End time is 3 days for all examples. Initial and final conditions are: (a) $(\phi_0,\lambda_0)=(\pi/4,0)$,
$(\dot{\phi}_0,\dot{\lambda}_0,\dot{\psi}_0)=(0,0,10\Omega)$, $(\phi_f,\lambda_f)= (\pi/4,0.1)$,
$(\dot{\phi}_f,\dot{\lambda}_f,\dot{\psi}_f)=(0,0,10\Omega)$; (b) $(\phi_0,\lambda_0)=(\pi/4,0)$,
$(\dot{\phi}_0,\dot{\lambda}_0,\dot{\psi}_0)=(0,0,\Omega)$, $(\phi_f,\lambda_f)= (\pi/4,\pi/3)$,
$(\dot{\phi}_f,\dot{\lambda}_f,\dot{\psi}_f)=(0,0,\Omega)$; (c) $(\phi_0,\lambda_0)=(\pi/4,0)$,
$(\dot{\phi}_0,\dot{\lambda}_0,\dot{\psi}_0)=(0,0,10\Omega)$, $(\phi_f,\lambda_f)= (\pi/4,\pi/3)$,
$(\dot{\phi}_f,\dot{\lambda}_f,\dot{\psi}_f)=(0,0,10\Omega)$.} \label{fig:controldisc}
\end{figure*}

\subsection{Implementation of the Optimization Problem} 

We employ the mid-point approximation to obtain the discrete
quantities in Section~\ref{sec:control}. The cost function is a measure of total control effort, defined by
\begin{equation} 
	J = \int_{t_0}^{t_f} (f_{\phi}^2 + f_{\lambda}^2) {\rm d} t, \quad J =
\int_{t_0}^{t_f} f_{\psi}^2 {\rm d} t, 
\end{equation}
for the particle and the spinning disc,
respectively. The left and right discrete forces are approximated by $f^-_n=f^+_n=\frac{4}{h}(f_n+f_{n+1})$ and the
discrete Lagrangian is given by
\begin{equation} L_d(q_n,q_{n+1}) =
hL\left(\frac{q_{n}+q_{n+1}}{2},\frac{q_{n+1}-q_{n}}{h}\right) . 
\end{equation}
The discrete cost
function for the particle is of the form 
\begin{equation} 
	J_d =
\sum^{N-1}_{n=0}\frac{h}{4}\left[(f_{\phi,n}+f_{\phi,n+1})^2 + (f_{\lambda, n}+f_{\lambda,n+1})^2\right], 
\end{equation}
while that for spinning disc is
\begin{equation} J_d =
\sum^{N-1}_{n=0}\frac{h}{4}(f_{\psi,n}+f_{\psi,n+1})^2. 
\end{equation}

The solution procedure is as follow: (i) guess the trajectory for a small number of discrete steps, often the geodesic
connecting starting and ending positions; (ii) solve the discrete optimization problem using sequential quadratic
programming (SQP) method to obtain optimized trajectory and cost, done in Matlab; (iii) use result in (ii) as a reference
in generating initial condition for more discrete steps, repeat (ii); (iv) compare the results, if the difference is small
enough, numerical solution is obtained, otherwise repeat (iii).

\subsection{Numerical Results of Motion Planning}

The results for particle motion under optimal control are shown in Figure~\ref{fig:controlparticle}. Initial and final
velocities are set to be zero for these simulations. Clearly, the optimal trajectories depend not only on the initial and
final conditions but also on the time period $t_f-t_0$. In Figures~\ref{fig:controlparticle}(a)
and~\ref{fig:controlparticle}(c), the time period is 1 day while the trajectories in \ref{fig:controlparticle}(b) and
\ref{fig:controlparticle}(d) correspond to the same initial and final conditions but with a time period equal to 3 days. In
order to arrive at final positions on time for \ref{fig:controlparticle}(b) and \ref{fig:controlparticle}(d), the particle
must travel further, while for \ref{fig:controlparticle}(a) and \ref{fig:controlparticle}(c) the trajectories are obviously
shorter. Yet the shorter trajectories come with the price of larger control efforts, $J_d|_a=66.2$, $J_d|_b=23.4$,
$J_d|_c=54.7$ and $J_d|_d=14.3,$ which can be interpreted physically as follows: although the particle in
\ref{fig:controlparticle}(b) and \ref{fig:controlparticle}(d) cover more distance, the fact that it has more time to reach
its final destination means that it can better exploit the rotation of the earth.

 The examples for a spinning disc under optimal control are shown in Figure~\ref{fig:controldisc}. The first row
illustrates the trajectories while the second row shows the corresponding spinning velocities $\dot{\psi}$. The initial and
final conditions are: $\phi_0=\phi_f=\pi/4$, $\lambda_0=0$, and $\lambda_f$ are chosen to be $0.1$, $\pi/3$ and $\pi/3$
respectively, also $\dot{\phi}_0=\dot{\lambda}_0=\dot{\phi}_f=\dot{\lambda}_f=0$, and $\dot{\psi}_0=\dot{\psi}_f$ are
chosen to be $10\Omega$, $\Omega$ and $10\Omega$ respectively. Since the control torque is only applied in $\psi$
direction, the disc is under-actuated and motion planning is harder than in the particle case. Comparing
\ref{fig:controldisc}(a) with \ref{fig:controldisc}(c), one can observe that the trajectories have very similar behavior in
the first half period. However, in the second half period, in order to travel further in \ref{fig:controldisc}(c),
$\dot{\psi}$ needs to be increased more than that in \ref{fig:controldisc}(a), i.e. the control effort needs to be larger.
Meanwhile, trajectories in \ref{fig:controldisc}(b) and \ref{fig:controldisc}(c) have similar behavior in the second half
period, yet in the first half, the trajectories are quite different because of the different initial conditions. It is
important to note that these examples result in net latitudinal changes. A net longitudinal motion seems not to be
achievable by applying a control torque in $\psi$ only. This issue will be investigated in future work.

\section{Summary} 
\label{sec:conc}

The motion planning for a particle and a spinning disc on earth was considered using an optimal control approach. These
models are relevant for understanding geophysical vortex motion as well as for controlling the motion of atmospheric
drifters such as high altitude balloons. In this paper, we employed an existing model of earth as a perfect sphere with an
additional potential, then derived the equations governing the motion of a particle and a spinning disc moving on earth
using Lagrangian mechanics, and showed solutions using direct numerical integration. We also examined the steady drift
obtained in~\cite{McDonald1998}, and gave conditions for its validity. Motion planning for the particle and spinning disc
was examined using discrete mechanics and optimal control. Rigorous proof of controllability of the spinning disc via one
control torque about its spin axis remains an open question to be addressed in future work. Future directions include
application of the earth model and the control methods used in this paper to the problem of multiple vortices on a sphere
discussed in~\cite{Newton2001}.


\end{document}